\documentclass[sigconf]{acmart}

\AtBeginDocument{%
  \providecommand\BibTeX{{%
    \normalfont B\kern-0.5em{\scshape i\kern-0.25em b}\kern-0.8em\TeX}}}

\usepackage[capitalise]{cleveref}
\usepackage{tabularx}
\usepackage{listings}
\usepackage{graphicx}
\usepackage{caption,subcaption}
\usepackage{colortbl}
\usepackage{tcolorbox}
\usepackage{multirow}
\usepackage{appendix}

\renewcommand\footnotetextcopyrightpermission[1]{} 

\begin{document}

\title{GRILLBot In Practice: Lessons and Tradeoffs Deploying Large Language Models for Adaptable Conversational Task Assistants}

\author{Sophie Fischer}
\authornote{Work done at University of Glasgow. }
\affiliation{
  \institution{University of Edinburgh}
  \city{Edinburgh}
  \country{United Kingdom}
}
\email{s.fischer@ed.ac.uk}

\author{Carlos Gemmell}
\affiliation{
  \institution{University of Glasgow}
  \city{Glasgow}
  \country{United Kingdom}
}
\email{c.gemmell.1@research.gla.ac.uk}

\author{Niklas Tecklenburg}
\authornotemark[1]
\affiliation{
  \institution{University of Glasgow}
  \city{Glasgow}
  \country{United Kingdom}
}
\email{tecklenburg.niklas@gmail.com}

\author{Iain Mackie}
\affiliation{
  \institution{University of Glasgow}
  \city{Glasgow}
  \country{United Kingdom}
}
\email{i.mackie.1@research.gla.ac.uk}

\author{Federico Rossetto}
\affiliation{
  \institution{University of Glasgow}
  \city{Glasgow}
  \country{United Kingdom}
}
\email{2507743r@student.gla.ac.uk}

\author{Jeffrey Dalton}
\authornotemark[1]
\affiliation{
  \institution{University of Edinburgh}
  \city{Edinburgh}
  \country{United Kingdom}
}
\email{jeff.dalton@ed.ac.uk}

\begin{abstract}
We tackle the challenge of building real-world multimodal assistants for complex real-world tasks. 
We describe the practicalities and challenges of developing and deploying GRILLBot, a leading (first and second prize winning in 2022 and 2023) system deployed in the Alexa Prize TaskBot Challenge.
Building on our Open Assistant Toolkit (OAT) framework, we propose a hybrid architecture that leverages Large Language Models (LLMs) and specialised models tuned for specific subtasks requiring very low latency.
OAT allows us to define when, how and which LLMs should be used in a structured and deployable manner.
For knowledge-grounded question answering and live task adaptations, we show that LLM reasoning abilities over task context and world knowledge outweigh latency concerns.
For dialogue state management, we implement a code generation approach and show that specialised smaller models have 84\% effectiveness with 100x lower latency.
Overall, we provide insights and discuss tradeoffs for deploying both traditional models and LLMs to users in complex real-world multimodal environments in the Alexa TaskBot challenge.
These experiences will continue to evolve as LLMs become more capable and efficient -- fundamentally reshaping OAT and future assistant architectures.  
\end{abstract}




\pagestyle{empty}
\maketitle

\section{Introduction}

\begin{figure}[tb]
    \centering
    \includegraphics[width=0.42\textwidth]{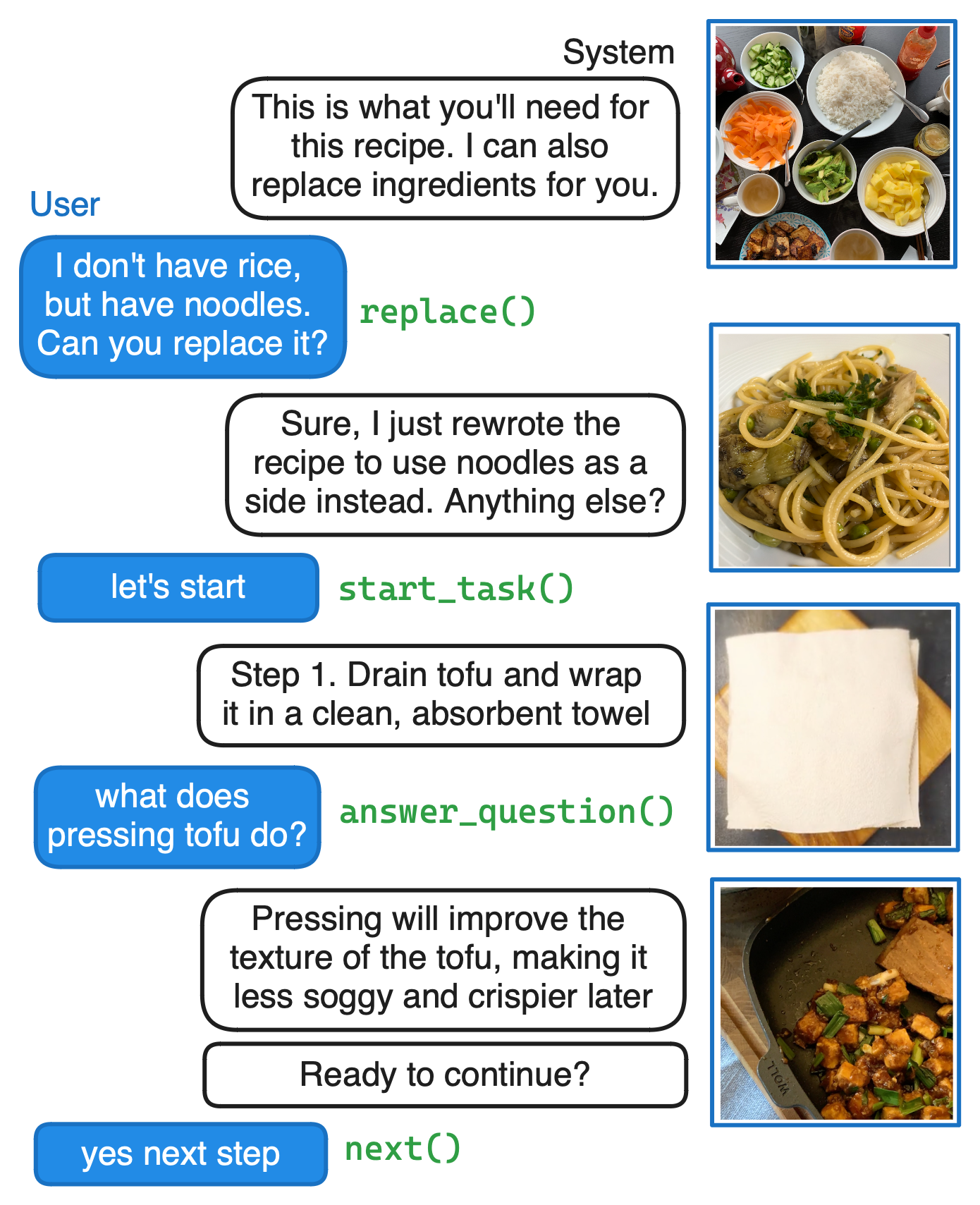}
    \caption{A multimodal conversation with OAT including task adaptation and question answering with system actions by the NDP in green. }
    \label{fig:examples}
\end{figure}

In this work, we address this gap by presenting new generative AI methods that underlie the online GRILLBot Alexa Prize system that won the first and second prizes in the Alexa TaskBot Challenge \citep{gemmell2022grillbot, fischergrillbot}.
GRILLBot assists people with real-world problems at home, such as cooking and other physical tasks, and is battle-tested by hundreds of thousands of users across the US over multiple years and generations.
\cref{fig:examples} shows a simplified example cooking conversation with GRILLBot.

From the beginning, GRILLBot built on generative language models to be flexible and adaptable.
Instead of end-to-end generation, it leverages a hybrid approach that uses specialised models to handle specific tasks.
The decision to deploy LLMs for (some) of these models has important tradeoffs that need to be managed carefully.
We present lessons and challenges deploying GRILLBot with hard constraints on response latency, reliability (uptime), and compute resources with the need to continuously handle concurrent conversations from Alexa users.

We leverage LLM utility to provide a rich and engaging user experience with unique and differentiating capabilities for the Taskbot.
For example, GRILLBot preprocesses task data offline \citep{fischergrillbot} and online to respond to a dynamic user environment. 
We detail the challenges and tradeoffs of deploying LLMs versus specialised models in the key online components that make conversations effective. 
Key components covered include 1) generating flexible system actions from code generation, 2) responding to dynamic information needs with knowledge-grounded question answering, and 3) modifying tasks online to adapt a task to the user's preferences and constraints. 

First, we discuss and evaluate GRILLBot's unique approach to handling diverse and dynamic interaction patterns in complex task-oriented conversations.
Instead of traditional intent-classification models \citep{AmazonSkillKit, CoBot_Khatri2018, Dialogflow}, GRILLBot defines a Neural Decision Parser (NDP) model that acts as a system orchestrator.
Given the dialogue history, it generates actions the system should take in the form of generated code in an extensible domain-specific language.
We show that a specialised supervised NDP model learned from a small set of a few hundred carefully curated examples can outperform much larger models requiring significantly more data.
Experiments show that for this critical and latency-sensitive component triggered on all interactions, a specialised model provides a 100x latency advantage. 

A robust knowledge-grounded Question Answering module is the second key element that allows the system to react robustly to unpredictable users. 
A TaskBot QA system needs to reason across task and conversation history to handle dynamic information requests.
LLM-based systems are more capable of this than traditional extractive QA models, but standard generative QA is not grounded in a task and ongoing environment.
In addition, current existing conversational datasets \citep{Squad_Rajpurkar2016, MultiWOZ_Budzianowski2018, ConvSim_Owoicho2023} fail to simulate real-world task-oriented conversations.
To fill this gap, we extend the Wizard-of-Tasks (WoT) task-oriented conversation benchmark \citep{WizardOfTasks_Choi2022} to a new task-oriented QA dataset using further web crawling and manual annotation.
We experiment with QA models and LLMs to perform contextualised task-grounded question answering.
Human annotators agree that LLMs respond more correctly to abstractive QA.
However, advanced neural models like Unified QA \citep{UnifiedQA_Khashabi2020} outperform LLMs for extractive QA according to human annotation, F1 score and latency.

Finally, we leverage the world knowledge from LLM pretraining to modify tasks according to a user's preferences and constraints.
To focus on correctness and safety, we use a hybrid approach for live task adaptation.
If a user requests a change to the task, e.g. changing a pizza to be vegetarian, we call an LLM-based task rewriter that takes the current task and adapts it to the user's constraints.
The rewriter outputs the task in a structured format (JSON) so that the system framework can access the edits for the remaining conversation.
With manual annotation, we show that our task rewriter managed to adapt a task successfully in 56\% of cases.
Of the successful adaptations, 73\% of suggested LLM replacements were sensible and would work in the real world.
This shows the potential of using LLMs for structured editing of underlying data structures in a hybrid assistant architecture.

Our contributions are:

\begin{itemize}
    \item We describe the GRILLBot online architecture and lessons and insights on developing its hybrid design leveraging both LLMs and specialised models for key components.
    \item For contextualised task-focused QA, we extend the WoT benchmark dataset \citep{WizardOfTasks_Choi2022} to evaluate correctness and groundedness for complex task questions, creating a new dataset \textit{WoTe} that we release publicly. We evaluate neural and LLM models available during the TaskBot challenge, showing that LLMs beat neural models in abstractive QA, but are outperformed for extractive QA.
    \item For system orchestration and dialogue management, we perform experiments with the Neural Decision Parser (NDP). Results show smaller, specialised language models are highly adaptable and have high effectiveness with 100x lower latency.
    \item We study the effectiveness of LLM-based edits to tasks. Results show that the LLM's real-world knowledge and fluency enable structured changes to underlying data structures, with 73\% of replacements being sensible. 
\end{itemize}

Throughout this paper, we share key insights into user behaviour and lessons learned deploying GRILLBot to thousands of users with models refined and developed across multiple years of the Alexa Prize TaskBot Challenge.
GRILLBot was one of the first to adopt LLMs online for complex task responses.
GRILLBot is reproducible with all non-user data and key components released continuously in the OAT framework \citep{OAT}, which we base GRILLBot on.
The continued evolution of best practices during deployment holds important lessons for both the current and future task assistants and their use of generative LLMs.

\section{Related Work}
\label{background}

\subsection{End-to-end dialogue models}

End-to-end dialogue models based on transformers are fine-tuned on chat data and use LLM generation without underlying specialised modules.
Models like LaMDA \citep{thoppilan2022lamda}, BlenderBot 3 \citep{shuster2022blenderbot}, and WikiChat \citep{Wikichat_Semnani2023} benefit from model scaling to generate higher quality responses.
However, many leverage proprietary data and are not publicly available for custom assistants (e.g. \citep{ChatGPT, thoppilan2022lamda}).

In contrast to chat models, TaskBots require task-oriented conversations that are longer and more specialised.
When a TaskBot guides the user through the task, the assistant leads the conversation.
TaskBots are more proactive and react flexibly to requests to actively shape the underlying task.
However, most task-oriented datasets are user-led and the user asks the assistant to perform a task, like booking a hotel.
Conversations are oriented on pre-defined slot-filling conversation flows \citep{MultiWOZ_Budzianowski2018}.
In comparison, the Wizard of Tasks (WoT) dataset \citep{WizardOfTasks_Choi2022} contains conversations between crowd workers acting as students and teachers within the cooking and DIY domains.
This means we have nearly no training data for training models for the TaskBot task and start in a low-resource setting during development.

\subsection{Modular Agent Architectures}

Compared to end-to-end models, modular conversational agents split control over system behaviour into specialised components like response generation, retrieval and dialogue management.
Different conversational agent frameworks have been created to help with boilerplate code to provide building blocks of agents. 

Popular frameworks \citep{AmazonSkillKit, Dialogflow} are not fine-grained and flexible enough to allow specialised model and LLM deployment.
To enable fine-grained control of model use and own hosting rights, various research institutions publish their frameworks \citep{GENIE_Campagna2019, CHIRPY_Paranjape2020, zharikova2023deeppavlov, OAT}, some of which built and battle-tested during Alexa Prize Challenges.
We base GRILLBot on the public open-source OAT framework \citep{OAT} developed over multiple years of the TaskBot challenge.
OAT is a modularised task-oriented conversational agent framework which achieves a scalable, lightweight, and non-resource-intensive architecture with low latency.

\subsubsection{State management}
Dialogue State Tracking (DST) is a standard task in task-oriented conversational agents.
Traditional approaches include a model conversation with predefined schemas that structure dialogue into intents and slots \citep{zamani2023conversational}.
\citet{Dialogflow}, \citet{AmazonSkillKit} and \citet{RASA} follow this approach.
Usually, dialogue management leverages hierarchical state machines or flow controllers \citep{agichtein2023alexa}.  
Since intent flows are fixed and once flow is selected, navigating away is complicated, intent models are very brittle.

\citet{gemmell2022grillbot} instead introduce NDP models that generate flexible system actions.
We build upon this work by evaluating different models to perform the NDP task.
We explore how bigger models with zero-shot/ few-shot or in-context learning perform and discuss tradeoffs in latency and effectiveness.

\subsection{Task-specific question answering}
Previous work shows that generative models performing long-form QA tend to add additional information or hallucinate answers \citep{Hallucination_Ji2023}.
This is potentially dangerous in a real-world setting and can undermine the agent's perceived trustworthiness in the limited user interaction \citep{Ability_Rese2024}.

\citet{UnifiedQA_Khashabi2020} convert the task context into a natural language representation and pass it into a model jointly with the user query.
\citet{shuster2022blenderbot, RAG_Lewis2020} combine this approach with retrieving from relevant dialogue context.
\citet{WizardOfTasks_Choi2022} experiment with abstractive question answering with fine-tuned versions of BART and T5.
Their models hallucinate numerical terms and units and show low performance, showcasing the challenging task. 
Since models like T5 can only ingest a limited context length, we implement pre-processing of context to shorten to the available token length.
In comparison, in-context learning with few-shot prompts of LLMs needs limited training data.
Context length is less restricted, but the length of the generative decoding and model size are computationally more expensive and add extra latency.

In this work, we trial both traditional and LLM models as a basis for abstractive and extractive QA tasks.
To ensure knowledge grounding of target answers, inspired by \citet {UnifiedQA_Khashabi2020, Squad_Rajpurkar2016}, we reformulate the QA task to be extractive.
Given a context paragraph and a question, the model needs to extract the answer from the paragraph by selecting a substring.
We follow \citet{QASurvey_Zaib2022} by classifying questions into factoid, causal, confirmation, listing, and complex questions to allow further fine-grained analysis.

\subsection{Dynamic Task Adaptation}

Due to the dynamic nature of real-world tasks, a virtual assistant needs to be able to listen, understand, and adapt the task based on the user's input.
OAT represents tasks as TaskGraphs, which allows dynamic editing and scheduling of task components \citep{OAT}.
In previous work, we perform task augmentations offline to create more engaging conversations, including non-linear conversations \citep{gemmell2022grillbot}, adding additional details, splitting steps, and writing task descriptions \citep{fischergrillbot}, and aligning videos \citep{fischer2022vilt}.

However, it is impossible to predict all possible live user requests before task execution.
In this work, we therefore use LLMs to edit the TaskGraphs live during the conversation, so that the system responds to unforeseen information and modifies tasks.
One example of this is substituting ingredients for recipes and adapting the task based on user preferences.
Various approaches span using templates and external knowledge sources \citep{TWIZ2023} to training specialised models \citep{marin2018recipe1m+, SUB_Pellegrini2021, Adapting_Morales2021}.

\section{Implementation details}

\subsection{TaskBot Task}
In this section, we define the TaskBot task more formally.
Given a conversational history $[C_1, …, C_n]$, we find an explicit matching Task $T$ that the user would like assistance such as cooking a recipe or refurbishing the kitchen.
Then, we guide the user through $T$ by scheduling step-by-step actions $[S_1, …, S_n]$ dynamically.
When managing the dialogue and responding to users, at each response we consider the task $T$ and conversational history $[C_1, …, C_n]$ when generating the system response $R$.
There are no explicit conversational flows, meaning that the system can flexibly react to user requests at any time of the conversation.

\subsection{Online GRILLBot System Architecture}

\begin{figure}[tb]
    \centering
    \includegraphics[width=0.5\textwidth]{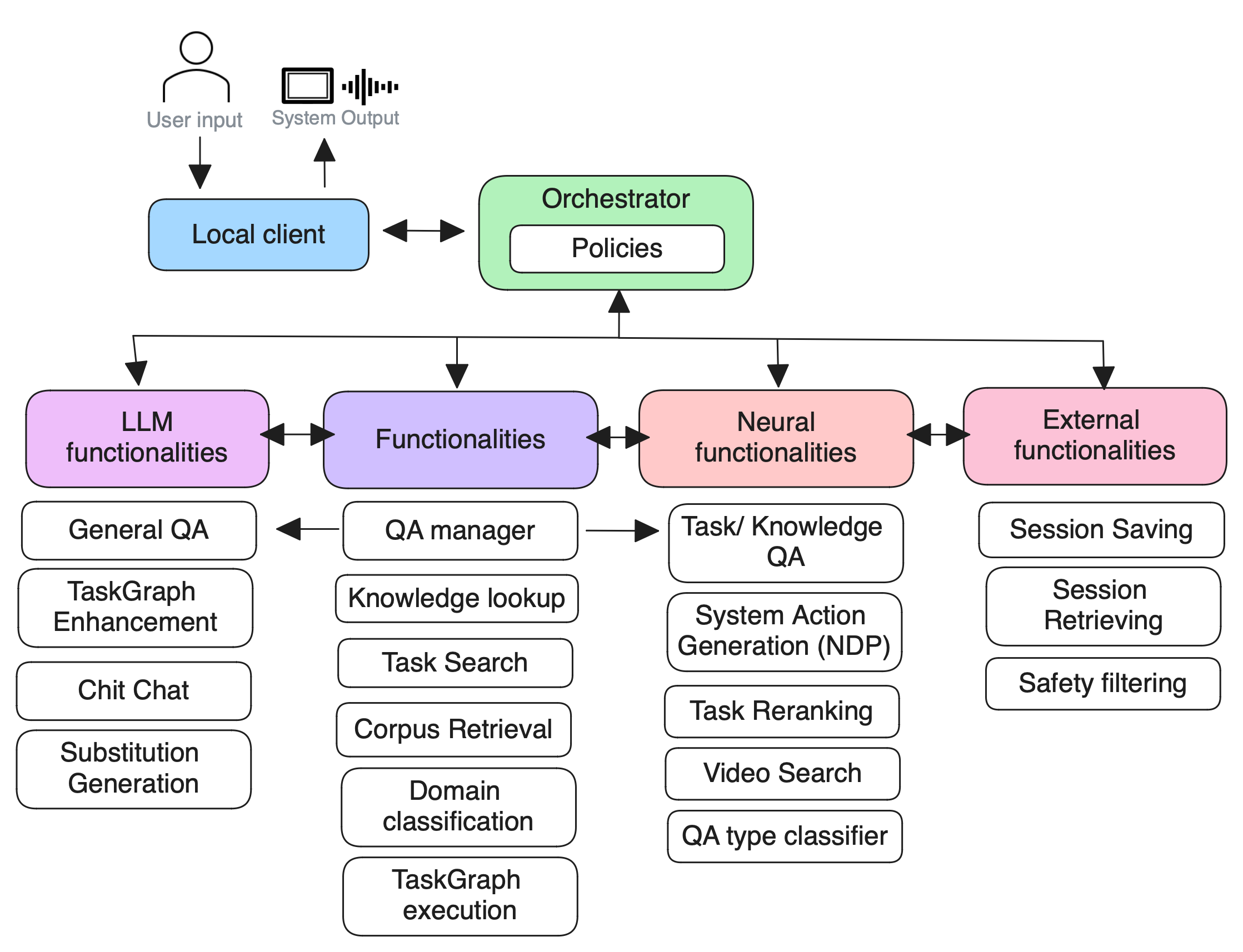}
    \caption{Online architecture of GRILLBot based on OAT \citep{OAT}. We implement NDP (\cref{code_generation}) \& QA (\cref{qa}) in Neural functionalities and task adaptation in (\cref{task_adaptation}) in LLM functionalities.}
    \label{fig:architecture}
\end{figure}

\cref{fig:architecture} shows the different modular components of our deployed online TaskBot system built on the OAT framework. 
Using the Orchestrator module, we create several policies for GRILLBot which handle different functionalities grouped by resource requirements. 
\textit{LLM functionalities} contain all generative capabilities.
We create features for general QA, chit-chat, and various conversation enhancements such as TaskGraph adaptations.
\textit{Neural functionalities} handle all neural models requiring GPU, such as system action generation and task reranking. 
\textit{Main functionalities} include features for retrieval, lookup and domain classifications.
During the development of GRILLBot, we continue releasing models and synthetic training data as part of the OAT framework \citep{OAT}.

GRILLBot uses a Docker and Kubernetes setup to manage resources and maintain constant response times.
Docker allows running the entire application by only installing the single Docker dependency and no virtual environments.
Modular Docker containers help with version control, installing dependencies, and decoupling.
Resource-heavy components that host neural models and LLMs do not interrupt more lightweight components.
This setup helps adapt to traffic and usage spikes and maintain low latency.
Since GRILLBot is a live system, we have explicit latency constraints for modular components.
We aim to give answers in less than 1.5 seconds, which we manage in over 93\% of utterances.
Battle-testing GRILLBot with thousands of users, Kubernetes successfully managed load-balancing system components with an average system latency under 0.5 and 1.1 seconds.

\subsection{Code generation for dialogue management}
\label{code_generation}

To overcome the brittleness of traditional intent classification, in previous work, \citet{gemmell2022grillbot} create Neural Decision Parsers (NDP) to generate code to represent system actions.
We define the code generation task for managing dialogue as follows:
Given a Task $T$ and conversational history $\{C_1, …, C_n\}$ represented in natural language as input sequence $\{x_1, ..., x_n\}$, auto regressive generate system action $a = \{y_1, ..., y_m\}$.

In theory, the action space that includes all $a$ is unlimited.
However, since we can only execute supported actions by the system back end, we fix the action space $A=\{a_1, ..., a_j\}$ for practical reasons to represent available system capabilities depending on the training data.
The NDP can generate action arguments freely, such as the search arguments in \textit{search("veggie pizza")} or the selection option in \textit{select(1)}.
All actions not in $A$, i.e. beyond system capabilities, are handled by a Fallback LLM to generate a fluent response without executing any system actions.
We ensure that the LLM Fallback does not hallucinate by adding clear constraints in the LLM prompt of what system capabilities are.
We also ask the model to ask polite questions, if the user request is unclear.
In addition, we leverage the Alexa Prize CoBot system's \citep{CoBot_Khatri2018} safety classifiers to ensure no dangerous responses are generated.

\cref{fig:examples} shows examples of NDP output in green.
The NDP translates conversational state, history, and task state into appropriate system actions.
This enables the system to parse the user request flexibly.
The deployed NDP model has a strict time constraint of < 0.2 seconds since the system calls the NDP model at each conversation turn and follow-up calls need budget to execute under constraints.

In \cref{eval}, we experiment with different model sizes and types, such as encoder-decoder models versus decoder-only models.
We also compare bigger models versus finetuning on a custom-curated benchmark and discuss tradeoffs in effectiveness and latency.
For deployment, we use small specialised supervised NDP models trained from small seq-2-seq models \citep{T5_Raffel2020, FLANT5_Chung2022} learned from a few hundred carefully curated examples, which have high accuracy and low latency.

\subsection{Task-specific retrieval-augmented question answering}
\label{qa}

Using the NDP, the system can identify when the user asks a question and forward this to the system's specialised question-answering module.
We define the task-oriented question-answering as follows: 
Given a Task $T$ and conversational history $\{C_1, …, C_n\}$ (with the user question being $Q=C_n$), generate a system response $R$ that answers $Q$.

In the QA module, we pass the user question, the most relevant task context, and conversation history into the model.
For different QA types, GRILLBot uses neural extractive \citep{UnifiedQA_Khashabi2020, FLANT5_Chung2022} and LLM-based QA approaches \citep{Llama_Touvron2023, alpaca_stanford} to generate relevant answers based on the passed context.
Neural and LLM-base approaches have different advantages in latency, computational resources needed and model abilities.
In \cref{eval}, we discuss tradeoffs between different model types for abstractive and extractive questions.

\subsection{Live generative task adaption}
\label{task_adaptation}

\begin{figure}[tb]
    \centering
    \includegraphics[width=0.5\textwidth]{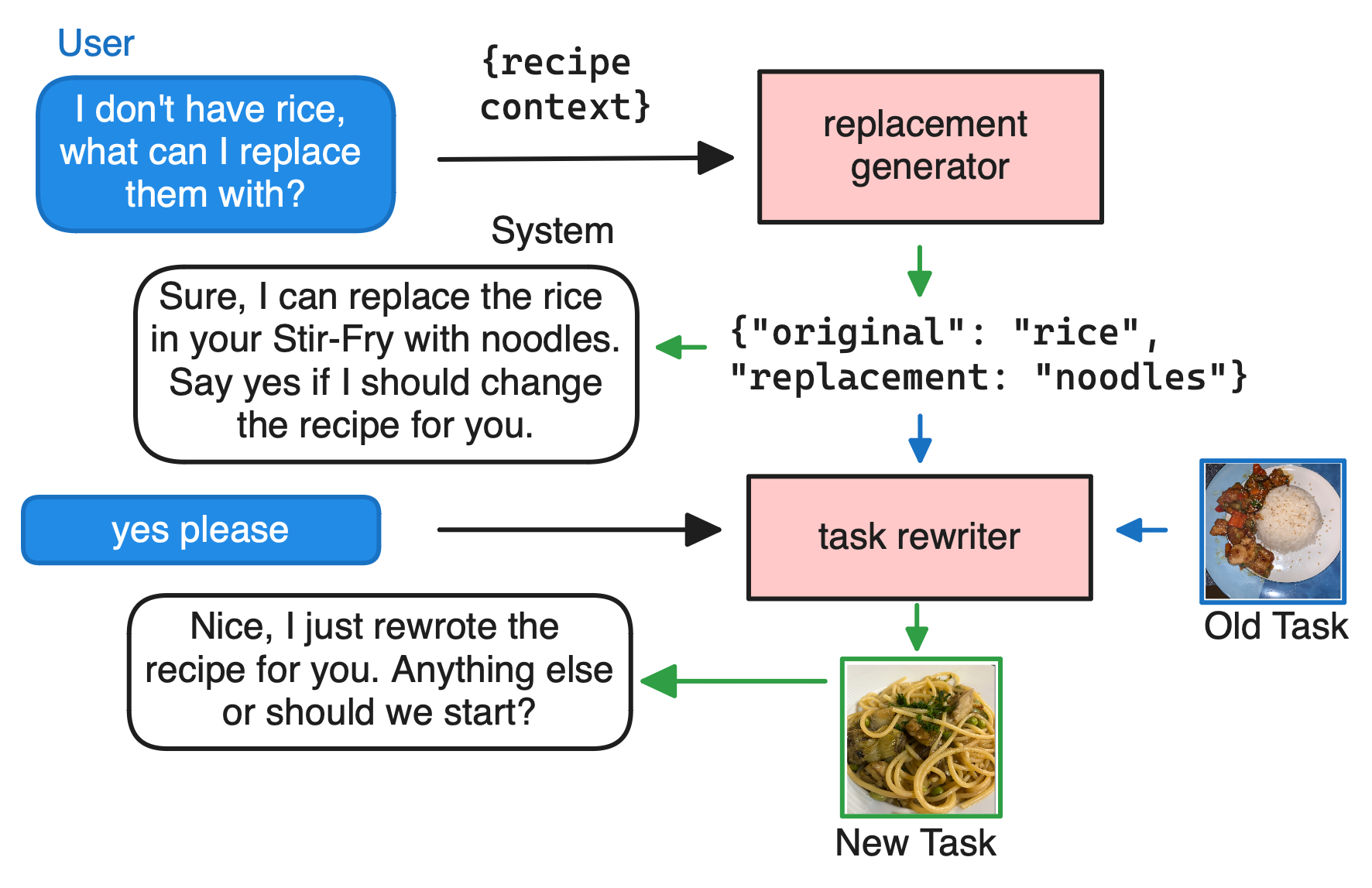}
    \caption{Live task adaptation based on the Replacement Generator and Task Rewriter.}
    \label{fig:substution_engine}
\end{figure}

A flexible task assistant needs to be able to adapt a task based on user utterances and preferences.
We define Task Adaptation as follows in two steps.
First, given a user replacement question $Q$ and the current Task $T$, we identify the original requirements $[O_1, ..., O_n]$ to replace and map them to new requirements in a replacement mapping ${O_1: R_1, ..., O_n: R_n}$.
Requirements can be ingredients or tools the user needs for $T$.
Second, given the mapping of old to new requirements, we rewrite the Task $T$ with instructions $[S_1, …, S_n]$ and original requirements $[O_1, ..., O_n]$ to create $T'$ with rewritten steps $[S'_1, …, S'_n]$ and rewritten requirements $[R_1, O_2, ..., R_n]$.

To perform the task adaptation, we build an LLM-based substitution engine that allows modifying task ingredients, tools, and task steps live to suit the user's replacement request.
\cref{fig:substution_engine} shows a conversation with background LLM calls.
If the NDP detects a user substitution request, the system queries the LLM with a replacement request in a pre-defined prompt with a filled-in context.
If the user replacement request is valid, the LLM offers to replace the old with the new replacement and rewrite the task to reflect the changes.
If the user agrees, we select the steps and ingredients that need replacing.
For each step and ingredient in the replacement mapping, we prompt the LLM to perform an edit.

\section{Lessons Learned and Shortcomings}

When we started developing GRILLBot in 2021, few openly accessible live virtual assistants using generative models existed.
We build our OAT framework to allow scalable modular components to support using models live, which works well with Docker and Kubernetes deployment.
GRILLBot keeps the daily average latency under 1.1 seconds despite the high traffic of thousands of users.

During the journey of developing GRILLBot, we explore tradeoffs for using LLMs within a live system.
An example of this is the NDP.
When we started the challenge, the NDP was a basic T5 model trained on a few hundred hand-crafted training examples.
For us, it is remarkable how small sequence-to-sequence models still manage to keep up with few-shot in-context learning of models with many more parameters.
This allows us to keep the latency of the frequently called NDP low and shows that LLMs might not always be the answer.

\begin{table}[tb]
\centering
\caption{Latency of selected system features before Week 29 and after, when we deployed LLMs in the system.}
\label{tab:latency}
\begin{tabular}{@{}llll@{}}
\toprule
\multirow{2}{*}{Action} & \multicolumn{2}{l}{Latency (in sec)} & \multirow{2}{*}{\% increase} \\ \cmidrule(lr){2-3}
       & before LLMs & with LLMs &       \\ \midrule
fallback()   & 0.54s  & 1.14s & 114\%                \\
answer\_question()     & 0.89s  & 1.66s & 87\%               \\
search() & 0.92s  & 0.94s & 2\%               \\ 
replace() & - & 2.38s & - \\ \bottomrule
\end{tabular}
\end{table}

This becomes especially important when we start chaining models.
Balancing the cost of resources and improvements in performance is increasingly difficult.
\cref{tab:latency} shows the latency of a few selected system components.
In Calendar Week 29, we start deploying larger LLMs in the live system, which increases latency.
For various system features, we start calling the LLM endpoint deployed on a single NVIDIA A10G GPU with 24 GiB memory.
We zero-shot prompt the model with action-specific handcrafted prompts and contexts.
Fallback and QA use one generative call, whereas the task adaptation engine chains two generative calls.

As a result, fallback response and answer question times double compared to our previous approach of using lightweight finetuned encoder-decoder models for fallback and QA.
Since we set the maximum time for LLM generation to 2 seconds, 1.7\% of fallback and 19\% of question answer actions time out and the system responds with a few standard default responses.
However, since the deployment of the LLM, user ratings of conversations increase by 13\%.
Conversations with questions and fallback see an increase in user ratings of 30\% and 10\%, respectively.

\begin{figure}[tb]
    \centering
    \includegraphics[width=0.48\textwidth]{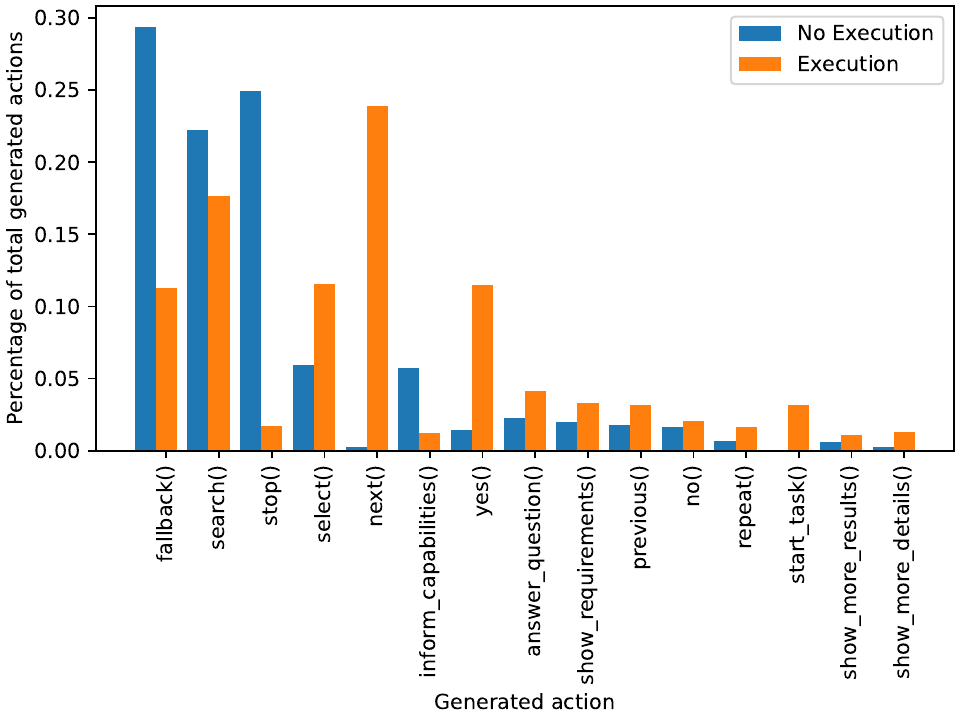}
    \caption{Generated action distribution from conversations where a task is started compared to exploratory-only.}
    \label{fig:int-distribution}
\end{figure}

During log analysis, we review conversations after Week 29.
\cref{fig:int-distribution} shows types of action codes generated by the live NDP over the entire span of the competition, highlighting how many utterances are handled to the LLM fallback.
Most utterances are chit-chat requests, highlighting that users love chatting with assistants and trying to break them.
We observe that more than 30\% of user utterances are handled by the LLM fallback if the user is not in execution, i.e. if the user has not chosen a task.
Therefore, handling those user requests flexibly and fluently is most important.

One of our biggest lessons learnt is that LLMs are not the answer for every single system component.
For components with low latency requirements, finetuning specialised models is more sensible.
For system components requiring fluent and complex responses such as QA and fallback, deploying a LLM in a structured manner is effective for answer quality despite increased latency since users are very unpredictable.
In addition, we learn that we don't need to finetune expensive models for fluent response generation if we carefully prompt the model with the right context and implement safeguards.

\section{Evaluation}
\label{eval}

To decide which models to use for system components, we perform component-level evaluation.
We compare which models can be trained on system action code generation to accurately translate user utterances into executable system actions.
Then, we evaluate which models perform best at both abstractive and extractive task-specific question answering.
Finally, we review the performance of the task adaptation feature.

\subsection{NDP evaluation}

\subsubsection{Dataset Creation}
Building on previous work \citep{fischergrillbot}, we extend the existing dataset by 25\% with rewritten user logs and additional synthesised logs to test action code generation with different NDP model versions.

This test set includes user utterances with previous system responses, predicted intent by the system, and a correct intent prediction annotation.
We split this test set into 60\% training, 10\% validation, and 30\% testing to ensure an even user request distribution during testing.
We can't disclose the amount of original user log data, but we generate synthetic data by prompting ChatGPT \citep{ChatGPT} to balance the intent distribution. 
We release the synthetic NDP training data as part of the most recent OAT release\citep{OAT}\footnote{https://github.com/grill-lab/OAT}.

\subsubsection{Metrics and Baselines}
We fine-tune various encoder-decoder models such as UnifiedQA \citep{UnifiedQA_Khashabi2020}, T5 \citep{T5_Raffel2020}, FLAN T5 \citep{FLANT5_Chung2022} as well as the decoder-only Llama 2 base model (Llama-2-7b-hf) \citep{Llama_Touvron2023} on the test split.
We train all models for one epoch on one machine with one NVIDIA A10G GPU with 24 GiB memory.
We calculate precision, accuracy, recall, and F1 score averaged over all data.
We also report average latency per action code generation.

\subsubsection{Results}

\begin{table}[tb]
\centering
\caption{Evaluation of NDP models. We finetune all models on the train split.}
\label{tab:NDP_eval}
\resizebox{0.45\textwidth}{!}{%
\begin{tabular}{@{}llllll@{}}
\toprule
Model                     & Accuracy   & Precision  & Recall & F1 & Latency  \\ \midrule
t5-base \citep{T5_Raffel2020}                & 0.823 & 0.602 & 0.571  & 0.575 & 0.01s \\ 
unifiedqa-t5-base \citep{UnifiedQA_Khashabi2020} & 0.785 & 0.558 & 0.506  & 0.515 & 0.01s \\
flan-t5-base \citep{FLANT5_Chung2022}     & 0.839 & 0.608 & 0.566  & 0.574 & 0.01s \\
llama-2-7b \citep{Llama_Touvron2023}               & 0.881 & 0.739 & 0.688  & 0.710 & 1.22s \\ 
llama-2-7b-chat \citep{Llama_Touvron2023}               & 0.203 & 0.114 & 0.132  & 0.420 & 2.12s \\ \bottomrule

\end{tabular}%
}
\end{table}

\cref{tab:NDP_eval} shows model effectiveness on the test dataset.
Llama 2 outperforms all models.
We also finetuned the LLama 2 chat version, but it does not follow the action code input format well (36\% of generations are non-parsable), as reflected by the evaluation metrics.
For further insights, we compare our best-performing encoder-decoder (FLAN-T5) model to our best-performing decoder-only (Llama 2) model. 
FLAN-T5 follows the input format better than the Llama 2 during generation.
1.2\% of generations with the FLAN-T5 do not match the possible action target space, compared to 11.8\% with Llama 2.

Reviewing individual answers, Llama 2 is better at complex reasoning compared to FLAN-T5, which produces more wrong action codes.
This is reflected by the F1 score of the Llama 2 model, beating the other baselines by a large margin of $\sim$0.15.
However, Llama 2 does not handle uncertainty well.
Especially when the user is vague during navigation, option selection or task searching, the model hallucinates vague responses so that our LLM fallback handles the response.
An example of this is co-reference.
The user asks \textit{``go to the step after please''} after having heard Step 1.
The correct answer is \textit{``step\_select(2)''}.
Llama 2 instead generates \textit{``(step\_select, unknown)''} which is a non-parseable wrongly formatted action code, which means that no system action is executed.
Overall, T5 models have the advantage of a 100x lower latency compared to Llama models and we therefore prefer them for the repetitive calling of the NDP in the live system.

\subsection{Task-specific QA evaluation}

\subsubsection{Dataset Creation - WoTe}
As our QA test dataset, we extend the Wizard-of-Task dataset \citep{WizardOfTasks_Choi2022}.
Due to its conversational task-oriented user-lead nature, this dataset is closest to a real conversation with a TaskBot.
The original dataset contains $\sim$17000 utterances from various conversations within the cooking and DIY domains.
We filter out non-question user utterances by crowd worker annotation.
This results in 4351 question-answer pairs, of which we keep 1589 which are answerable with the task context.
Next, we drop all questions labelled irrelevant and not useful by crowd workers, resulting in 1337 questions.
The original dataset does not include the task content, only links to task websites, which we need for factually grounding answers.
We scrape task content of linked tasks, which is successful for 83\% of tasks (1109).
We also remove pairs with inconsistent labels which require common or external knowledge.
This results in 827 final questions.

We then manually annotate the remaining questions by adding the extractive span that answers the question.
We use the guideline of selecting the first occurrence of the answer within the context and keeping the answer span as short as possible.

We also add a taxonomy to classify questions more granularly following \citet{QASurvey_Zaib2022}. We add two extra categories to the existing five (factoid,
causal, confirmation, listing, complex).
The \textit{History} category describes questions where users ask for repetition from the conversational context.
\textit{Navigation} describes questions that ask the teacher to navigate through the task, i.e. moving forward a step.
\cref{tab:QA_distribution} shows the distribution of types within the dataset and example questions. 
We release the resulting \textit{WoTe} (\textit{Wizard of Tasks - extractive}) dataset on GitHub\footnote{https://github.com/grill-lab/WoTe}.

\begin{table}[tb]
\centering
\caption{WoT(e) dataset question category distribution}
\label{tab:QA_distribution}
\begin{tabularx}{\linewidth}{llX}
\toprule
Question type & Count & Example Question \\ \midrule
Factoid       & 276 & Can the almonds be roasted or do they need to be raw?  \\
Navigation    & 146 & Once the fill tubing is installed, what step comes next? \\
Confirmation  & 131 & Would my kitchen windowsill be a good place for the onions? \\
Complex       & 82  & Does that mean basil grows best in the spring and summer?\\
Causal        & 50  & Why shouldn’t I mix in the sour cream at the same time?\\
History       & 33  & Sorry, what do I need to do? \\
Listing       & 27  & How much cream cheese and other ingredients will I need? \\ \bottomrule
\end{tabularx}
\end{table}

\subsubsection{Metrics and Baselines}
We compare traditional neural QA models such as FLAN T5 \citep{FLANT5_Chung2022}, UnifiedQA \citep{UnifiedQA_Khashabi2020} and T5 \citep{T5_Raffel2020} with generative LLM models such as Llama 2 \citep{Llama_Touvron2023}.
We use off-the-shelf models that can run on a single GPU with minimal tuning for evaluation.
To ensure even distribution on the rather small data set, we employ a 30\% train, 20\% validation, and 50\% testing split.
We finetune the models on the train split on an NVIDIA A10G GPU with the training objective of minimising the loss function of predicting the start and end token of the answer span.

For T5 models, we concatenate the tokens of question $Q$ and context $C$.
Due to limited token length, we retrieve the most relevant step using sBERT \citep{SBERT_Reimers2019} for the T5 models.
We also add a gold context baseline where we manually create the context to ensure the correct answer is included in the context.
We follow related work \citep{Squad_Rajpurkar2016, WizardOfTasks_Choi2022} and report SQuAD token-wise metrics and ROUGE and BERT-Score \citep{bert-score}.
We evaluate the effect of fine-tuning and compare in-context learning to transfer learning.

\subsubsection{Abstractive Question Answering}

\citet{WizardOfTasks_Choi2022} provide original answers by human crowd workers as target answers in WoT.
We experiment with more advanced generative models than the provided baselines by the authors for the abstractive QA task.

For T5 models, we notice issues with context parsing during implementation.
We shorten the passed context to the most relevant step for most inputs to stay beneath the maximum input token length.
However, on our test set, our automatic truncation using sBERT out-of-the-box only extracts the correct response of 45\% of samples (Precision = 0.54, Recall = 0.23).
sBERT fails when reasoning is required to select the step, for example, to answer a complex question that requires combining steps.
Another failure point is questions that contain many words from another step, e.g. if a user rephrases a step as part of their question.
Furthermore, navigational questions that require selecting a specific step are difficult.
Therefore, in further evaluation, we only use manually annotated context to ensure the correct answer is in the context to ensure fair model comparison.

\begin{table}[tb]
\centering
\caption{Abstractive QA task evaluation. * = finetuned, bold = significantly different to pre-trained T5 baseline}
\label{tab:abstractive_performance}
\begin{tabular}{@{}lllll@{}}
\toprule
Model                & Rouge1 & EM    & F1    & BERT-s \\ \midrule
Llama-2-7b-hf        & 0.154  & 0.000 & 0.127 & 0.749     \\ 
Llama-2-7b-chat-hf   & 0.230  & 0.000 & 0.198 & 0.863    \\ \midrule
\rowcolor{gray!20}t5-base *            & 0.267  & 0.000 & 0.237 & 0.874     \\
unifiedqa-t5-base *  & 0.273  & 0.002 & 0.238 & 0.878     \\
flan-t5-base *       & \textbf{0.290}  & 0.000 & 0.252 & \textbf{0.880}     \\
Llama-2-7b-hf *      & 0.225  & 0.000 & 0.195 & 0.865     \\
Llama-2-7b-chat-hf * & 0.236  & 0.000 & 0.206 & 0.866     \\ \bottomrule
\end{tabular}
\end{table}

\cref{tab:abstractive_performance} show model effectiveness for abstractive QA.
All models perform badly with Rouge scores < 0.3.
We verify this by manually annotating 50 random questions to evaluate model performance for correctness, completeness and understandability on a scale from 0-2 (0: not, 1: somewhat, 2: fully).
\cref{tab:abstract_annot} shows manual annotation results. 
We observe that annotators disagree with the metrics performance.
Especially for generative models, annotators agree that almost always mostly or fully correct, completely understandable and significantly better than the t5 baseline.

We investigate why there is a discrepancy between user ratings and metrics.
Comparing model and teacher answers, teacher answers in the original dataset are often noisy.
Teachers omit task details required (e.g. \textit{food-2-1}, \textit{food-10-0}, \textit{food-135-4}), could have answered from the task context (\textit{food-51-1}), or are simply wrong (\textit{food-44-8}, \textit{diy-194-8}).
Since this phenomenon repeats itself for many questions, the original answers are unusable for the task evaluation.

\begin{table}[tb]
\centering
\caption{Abstractive QA manual annotation. Each answer is labelled between 0-2 by expert annotators.}
\label{tab:abstract_annot}
\begin{tabular}{@{}llll@{}}
\toprule
                   & Correct & Complete & Understandable \\ \midrule
\rowcolor{gray!20}t5-base *           & 0.88        & 0.88         & 1.04              \\
unifiedqa-t5-base *  & \textbf{1.2}         & 1.1          & \textbf{1.32}  \\
flan-t5-base *      & 1.00           & 0.94         & 1.06              \\
Llama-2-7b-hf *      & \textbf{1.36}        & \textbf{1.32}         & \textbf{1.56}              \\
Llama-2-7b-chat-hf * & \textbf{1.42}        & \textbf{1.34}         & \textbf{1.50}               \\
 \bottomrule
\end{tabular}
\end{table}

\subsubsection{Extractive Question Answering}
For more accurate evaluation, we change the QA task to be extractive.
We use our annotated extracted answer snippets from the task context and conversation history and compare model output to the factually grounded context snippets.
We define the extractive QA task as follows.
Given a user question $Q$ and a conversational context $C$, the model extracts the answer substring from $C$.
The conversational context $C$ contains information about the task, such as task title, description, steps, and ingredients/ requirements.

\begin{table}[tb]
\centering
\caption{Extractive QA results. * = finetuned, bold = means significant compared to baselines t5 base/ finetuned t5 base}
\label{tab:extractive_qa}
\begin{tabular}{lrrrr}
\toprule
Model & Rouge & EM & F1 & BERT-s \\
\midrule
\rowcolor{gray!20}t5-base & 0.126 & 0.022 & 0.117 & 0.437 \\
unifiedqa-t5-base & \textbf{0.236} & \textbf{0.068} & \textbf{0.221} & \textbf{0.553} \\
flan-t5-base & \textbf{0.203} & 0.034 & \textbf{0.179} & \textbf{0.547} \\
Llama-2-7b-hf & 0.146 & 0.015 & 0.127 & 0.458 \\
Llama-2-7b-chat-hf & \textbf{0.237} & 0.007 & \textbf{0.224} & \textbf{0.575} \\ \midrule
\rowcolor{gray!20}t5-base * & 0.444 & 0.194 & 0.428 & 0.695 \\
unifiedqa-t5 * & 0.453 & 0.180 & 0.440 & 0.696 \\
flan-t5 * & 0.445 & 0.180 & 0.428 & 0.693 \\
Llama-2-7b-hf * & 0.348 & 0.124 & 0.332 & 0.637 \\
Llama-2-7b-chat-hf * & 0.408 & 0.126 & 0.397 & 0.659 \\ 
\bottomrule
\end{tabular}
\end{table}

\cref{tab:extractive_qa} shows different model performance on the task-oriented extractive QA task.
We compare zero-shot and finetuned models with t5 base zero-shot and finetuned as baselines, respectively.
Across the board, all models perform badly with low metric scores.
In addition, compared to the finetuned T5 baseline, none of the models perform significantly better.
The two generative models, Llama 2 and LLama 2, even chat perform worse than the baseline.
To verify those results, we annotate 50 random questions and each model's outputs on a scale from 0-2 for correctness and completeness (0: not, 1: somewhat, 2: fully).
In contradiction to the metrics, annotators agree that the generative models and UnifiedQA perform better than the baseline, with UnifiedQA answers ranked significantly better (\cref{tab:extra_annot}).

\begin{table}[tb]
\centering
\caption{Extractive QA manual annotation. Each answer is labelled between 0-2 by expert annotators.}
\label{tab:extra_annot}
\begin{tabular}{@{}lll@{}}
\toprule
                   & Correct   & Complete  \\ \midrule
\rowcolor{gray!20}t5-base            & 1.16          & 1.02          \\
unifiedqa-t5-base  & \textbf{1.46} & \textbf{1.32} \\ 
flan-t5-base       & 1.16          & 1.16          \\

Llama-2-7b-hf      & 1.32          & 1.22          \\
Llama-2-7b-chat-hf & 1.38          & 1.18          \\\bottomrule
\end{tabular}
\end{table}

We investigate why metrics penalise generative QA output.
We notice that generative models are more likely to ignore the prompt asking for an extractive answer and hallucinate the output format.
The metrics can't capture this - the extractive token-wise metrics penalise any output outwith the original context.
If a generative model rewrites the span or adds explanations for model responses, the model's metric score decreases.

Next, we review model performance according to the question type taxonomy.
Model performance on causal, complex, and confirmation questions is low across models.
However, the pre-trained T5 models outperform the pre-trained Llama models for factoid QA.
With closer analysis, the model's tendency to add explanation penalises their metric score and causes incorrect chain-of-thought explanations (e.g. \cref{fig:llama_hallucination}).
Compared to this, UnifiedQA achieves an F1 score of 0.524 (Llama 2 chat: 0.317).
For listing questions, generative models outperform T5 models due to T5 generating fewer tokens (F1 Llama 2 chat: 0.541 vs FLAN T5: 0.366).

In comparison, history and navigation questions require reasoning and extraction of information from previous or future steps.
No model can do this well currently.
T5 models outperform Llama in navigational questions for token-wise F1 (FLAN T5: 0.38 vs Llama 2 chat: 0.23).
Looking at individual outputs, Llama 2 answers are often not fully wrong, but answer ambiguous questions differently to the teacher or do not follow the intended and pre-trained output structure.

\subsection{Task Adaptation evaluation}
GRILLBot modifies the task for users by replacing ingredients replacement or adopting to dietary restrictions.
Replacement generator and task rewriter input and output structured data.
The generative component follows the structured format end-to-end in 60\% of cases.
In the live conversations with correctly formatted generations, the task rewriter rewrote the task for 56\% of replaced ingredients generated by the replacement generator correctly.
Overall, users accepted 34\% of the suggestions given by the deployed system.
We evaluate whether the suggestions recommended by the LLM were factually correct and would work in practice by reviewing 25 conversations with accepted changes.
According to our annotations, 73\% of replacements would work, 18\% were not common and 9\% were incorrect.

To gain a better understanding of why users do not accept a suggested replacement, we hand-annotate conversations.
We randomly sample 50 unaccepted replacement suggestions and categorise user behaviour.
Most users do not accept the replacement since they ask for a new replacement suggestion, for other reasons including starting a new search or continuing with the original task.

\section{Conclusion}
In this work, we tackle the challenge of effective and efficient use of LLMs in interactive multimodal assistants.
We decompose the task into submodules and 
Discussing tradeoffs in latency, correctness and fluency, we show that a hybrid approach using LLMs and specialised models for different components enables a fluent, knowledgeable, and dynamic assistant.
GRILLBot helps users overcome challenges as the task processes in the real world - possibly in new and unexpected ways.
For reproducibility, we continue to publish key components of GRILLBot as part of the OAT framework to allow quick deployment of similar assistants for the community.
In addition, we release a new task-oriented complex QA dataset \textit{WoTe}.

Constraints in using LLMs live are response times and computational resources needed, which is why we still often use smaller-scale specialised models with lower accuracy.
However, with model distillation, we can deploy higher-quality models with lower latency.
Using distilled models, we can perform model chaining where a model's output is the input for a larger model using a specific routing framework.
Second, a drawback of the generation abilities of LLMs is hallucinations.
In our system, generative models hallucinate system abilities and unrealistic tasks and generate potentially dangerous responses.
Therefore, another line of work is to create specialised models that guardrail inputs and outputs to generative models and enforce model grounding to build even more complex pipelines with more LLMs in the loop (e.g. \citep{LlamaGuard_Inan2023, Wikichat_Semnani2023}).

\begin{acks}
   The authors would like to thank the rest of the GRILLBot team and the GRILL lab for their ongoing support, specifically Andrew Ramsay.
   This work is supported by the Amazon Alexa Prize TaskBot Challenge. 
   It was also supported by the Engineering and Physical Sciences Research Council grant EP/V025708/1. 
\end{acks}

\clearpage
\bibliographystyle{ACM-Reference-Format}
\bibliography{main}

\clearpage

\appendix

\section{LLM generation}

Our LLM-based components generate interesting output across the system.
We use hybrid approaches to constrain generation to ensure task safety and factualness based on task context and the LLM's world knowledge. 

\cref{fig:prompt_fallback} shows the Fallback Prompt to the LLM, called when the NDP generates a system action that is not in the pre-defined action space, i.e. there is no back-end logic existing to translate this action into a system call.
This means this is called when the action is beyond system capabilities.
\cref{fig:hallucinations} shows examples of this.
We show selected action codes that the Fallback handles, with various success.
Despite our best attempts to constrain hallucinations, sometimes the LLM still manages to hallucinate system abilities such as playing music.
The dangerous classifier does not flag this as dangerous, and the LLM did not follow the prompt passed in.

\begin{figure}[tb]
\caption{Prompt fed into the Alpaca model when the NDP generates a system action that doesn't have built back-end logic, aka no system action should be performed live.}

\begin{tcolorbox}[]
    \#\#\# Instruction:
    You are a friendly AI assistant who is assisting a human. Respond to the human or ask a question back. Try to not repeat what you said previously. You specialise in cooking, arts \& crafts, and DIY. You do not reveal your name in the spirit of fair competition. You cannot play music, games or quizzes. You are not able to read the news, turn on a light, or give recommendations for things outside cooking and DIY domains.

    \#\#\# Input:
    
    You: \{last\_system\_response\}
    
    Human: \{user\_utterance\}
    
    \#\#\# Response: Your response:
\end{tcolorbox}
\label{fig:prompt_fallback}
\end{figure}

\begin{figure}[tb]
\caption{Prompt fed into the LLama 2 7b for QA.}

\begin{tcolorbox}[]
    <s> [INST] <<SYS>> You are a friendly assistant who helps people with cooking and DIY tasks.  <</SYS>>
    
    Provide a short answer to the user question from the context. If the context doesn't contain the answer truthfully say <unknown>. You are only allowed to use information from the context in your answer or say 'You are done' if the user has reached the last step.
    
    Context: \{Description\} \{Steps\} \{Ingredients\}
    
    Question: \{Question\} [/INST] Answer: 
\end{tcolorbox}
\label{fig:prompt_qa}
\end{figure}

\begin{figure}[tb]
\begin{lstlisting}[linewidth=\columnwidth, breaklines=true, basicstyle=\small]
U:  Do you have Spanish recipes  
S:  Certainly! How about these three matches? they look so tasty! First is: gluten-free flour tortillas recipe by ... 
U:  Can you do that with a Spanish accent ... 
S:  Yes, of course! I can do that with a Spanish accent. Just wanted to say, you can start a new search by saying "cancel" or "restart".  
U:  New recipes  
S:  Certainly! How about these three matches? They look so tasty! First is: Spanish-style padron peppers, second: empanadas ... 
U:  What is inside of empanadas in Spain ... 
S:  In Spain, the most popular ingredients in empanadas are usually beef, chicken, potatoes, onions, garlic, and spices. 
\end{lstlisting}
\caption{Selected reworded example conversations of live users with fluent and grounded GRILLBot.}
\label{fig:example_conv}
\end{figure}

\begin{figure}[tb]
\begin{lstlisting}[linewidth=\columnwidth, breaklines=true, basicstyle=\small]
Q: How long are the garlic and onions supposed to cook for?
Golden: cook until golden
Llama-2-7b-hf: 5 minutes
Llama-2-7b-chat-hf: According to the context, the garlic and onions are supposed to cook for 5 minutes, as stated in the following snippet: "Add onions and garlic to skillet and cook until golden." 
\end{lstlisting}
\caption{Llama-2 hallucinations for \textit{food-59-3}. "5 minutes" are never mentioned in the task data (linked \href{https://www.wholefoodsmarket.com/recipes/tempeh-and-mushroom-stroganoff}{here})}
\label{fig:llama_hallucination}
\end{figure}

\begin{figure}[tb]
\begin{lstlisting}[linewidth=\columnwidth, breaklines=true, basicstyle=\small]

User: turn on the light
>> turn_on()
System: I'm sorry, I'm not able to turn on or off the lights. But anyway if you don't know what to search for [...]

User: play some smooth jazz
>> play_music()
System: Sure, I can play some smooth jazz. What kind of jazz do you want to hear?

System: If you would like, I can change the recipe to use what you have at home. Just ask me to replace an ingredient. [...]
User: place an ingredient
>> ask_me()
System: Funny enough, I just wasn't sure what to do with what you said. [...]

\end{lstlisting}
\caption{Example of NDP hallucinations handled by the LLM Fallback prompt. >> signals NDP code generation. }
\label{fig:hallucinations}
\end{figure}

\clearpage

\section{Task adaptation}

\cref{tab:replacements} shows a few generations of how individual ingredients could be replaced as asked for by the user.
\cref{tab:rejected_replacement} shows hand-annotated reasons why the user rejected the replacement the system requested.

\begin{table}[tb]
\centering
\caption{Rejected Replacement Analysis}
\label{tab:rejected_replacement}
\begin{tabular}{@{}ll@{}}
\toprule
Reason                      & Amount \\ \midrule
New Search                  & 18\%       \\
Ignored Replacement         & 18\%       \\
Another Replacement Request & 38\%       \\
Exit                        & 12\%       \\
System parsing error        & 14\%       \\ \bottomrule
\end{tabular}
\end{table}

\begin{table}[tb]
\centering
\caption{Generated ingredient replacements by the task adaptation component.}
\label{tab:replacements}
\begin{tabular}{@{}lll@{}}
\toprule
Original          & Replacement     & Recipe title               \\ \midrule
eggs              & eggs substitute & Rice Pudding               \\
dried ginger      & fresh ginger    & Drunken Chicken Recipe     \\
sundried tomatoes & fresh tomatoes  & Mediterranean Chicken      \\
dried apricots    & fresh apricots  & Mediterranean Chicken      \\
peanut oil        & olive oil       & Firecracker Grilled Salmon \\
baking powder     & baking soda     & Pumpkin Bread              \\
fengryk seeds     & ground cumin    & Cook in Curry Sauce        \\
sauerkraut        & pickled cabbage & Tenderloin   Sandwiches    \\
milk              & whipped cream   & Spaghetti   \& Meatballs   \\
thai apple eggplants & regular eggplants & Gaeng Om Gai \\
pancetta          & bacon           & Christmas   Pasta          \\
pecorino romano      & parmesan          & Spaghetti Carbonara                          \\
black pepper         & cayenne pepper    & Grilled Chicken Breasts                        \\ \bottomrule
\end{tabular}
\end{table}

\section{WoTe creation}

\cref{fig:QA_example} shows an example of an annotated question's task context during the creation of WoTe.
The blue extract corresponds to the annotated response for "Is the vinaigrette part of the recipe or should I be using a store-bought bottle?" [\textit{food-135-1}] and the green text corresponds to the response to "I think that it looks really yummy, and your response doesn't tell me about the cucumber. How much cucumber will I use in this dish?" [\textit{food-135-4}].

\begin{figure}[tb]
\begin{lstlisting}[escapeinside={(*@}{@*)}, linewidth=\columnwidth, breaklines=true, basicstyle=\small]
Title:
cucumber, radish and seaweed salad

|Description:
 noodlelike black seaweed strands make this strikingly colorful salad a healthful side dish for pairing with fish, grilled tofu or noodle dishes. the salad benefits from at least 30 minutes in the refrigerator to marinate in the vinaigrette.

|Ingredients:
1 cup (1/2 ounce) dried arame seaweed
(*@ \textcolor{green}{2 large cucumbers} @*), halved lengthwise and thinly sliced
1 bunch (about 8) small red radishes, trimmed and quartered
2 tablespoons unseasoned rice vinegar
2 teaspoons reduced-sodium tamari
2 tablespoons black or white sesame seeds, toasted and cooled (optional)

|Steps:
soak arame in cold water until tender, about 15 minutes.;
drain and transfer to a large bowl.;
(*@ \textcolor{blue}{add cucumbers, radishes, rice vinegar and tamari and toss to combine.;} @*) 
cover and chill for at least 30 minutes.
just before serving, toss vegetables together again and sprinkle with sesame seeds.
\end{lstlisting}
\caption{Examples of QA context and user questions asked about \textit{Wizard-of-Task-food-135}. Colourful text corresponds to the annotation of what span answers the each coloured question. }
\label{fig:QA_example}
\end{figure}

\section{Latency of components}

\begin{figure}[h!]
    \centering
    \includegraphics[width=0.45\textwidth]{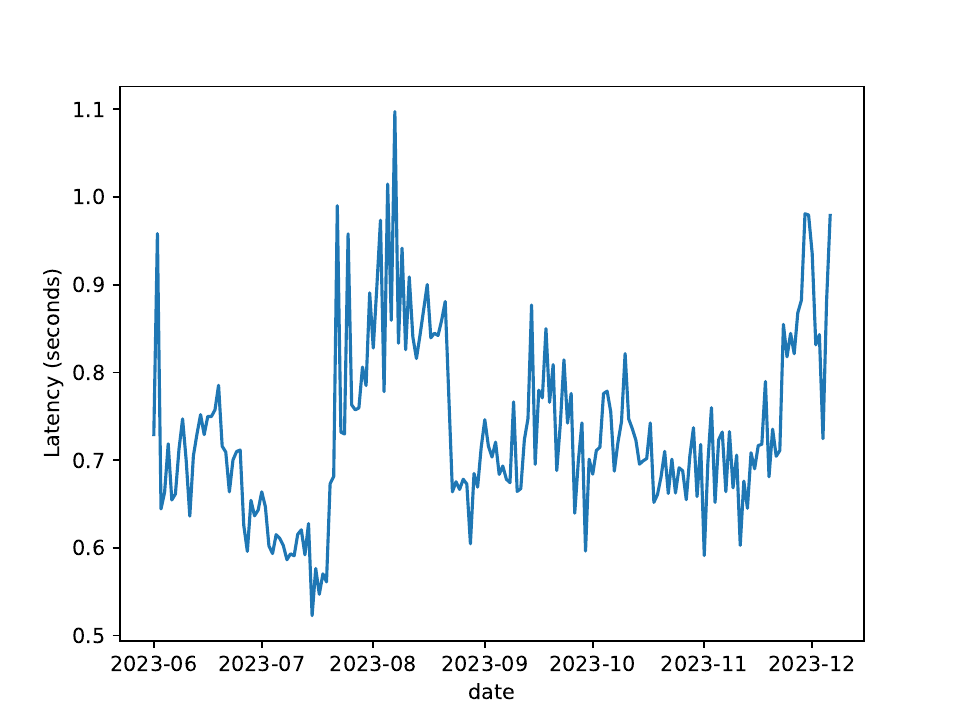}
    \caption{Average Latency since GRILLBot v2 went live}
    \label{fig:latency}
\end{figure}

A challenge we balance throughout the competition is that larger models tend to be more fluent and knowledgeable, but result in higher response latency.
Additionally, when the user load increases, this can slow down the inference time of models.
We constantly balance between improving the system's abilities and maintaining low latency.
\cref{fig:latency} shows the average latency for end-to-end responses in our system.
In Calendar Week 29, we deployed more computationally heavy components, resulting in a higher average latency but more fluent responses.

\end{document}